\documentclass[conference]{IEEEtran}
\IEEEoverridecommandlockouts
\usepackage{cite}
\usepackage{amsmath,amssymb,amsfonts}
\usepackage{algorithmic}
\usepackage{graphicx}
\usepackage{textcomp}
\usepackage{xcolor}
\usepackage{subfigure}
\usepackage[colorlinks,
            linkcolor=blue,
            anchorcolor=blue,
            citecolor=blue]{hyperref}
\usepackage{indentfirst}
\usepackage{mathrsfs}

\usepackage{multirow}
\usepackage[ruled]{algorithm2e}
\usepackage{multirow} 
\usepackage{bm}
\usepackage{verbatim}

\def\BibTeX{{\rm B\kern-.05em{\sc i\kerwn-.025em b}\kern-.08em
    T\kern-.1667em\lower.7ex\hbox{E}\kern-.125emX}}

\title{Max-Min Fairness for Uplink Rate-Splitting Multiple Access with Finite Blocklength}

\begin{document}
\author{

\IEEEauthorblockN{Jiawei~Xu, Yijie~Mao, \IEEEmembership{Member, IEEE}\vspace{-3.2em}}
\thanks{J. Xu is with Imperial College London. Y. Mao is with the School of Information Science and
Technology, ShanghaiTech University, Shanghai 201210, China (email: {j.xu20,b.clerckx}@imperial.ac.uk, maoyj@shanghaitech.edu.cn).}
}
\maketitle

\begin{abstract}
In this letter, we investigate the performance of Max Minimum Fairness (MMF) for uplink Rate-Splitting Multiple Access (RSMA) in short-packet communications. Specifically, considering a Single-Input
Single-Output (SISO) Multiple Access Channel (MAC), we optimize the transmit power allocation between the splitting user messages to maximize the minimum rate among users with Finite Blocklength (FBL) constraints. To tackle this problem, we propose a Successive Convex Approximation (SCA)-based approach. Additionally, we introduce a low-complexity scheme to design the decoding order at the receiver. Numerical results show that RSMA outperforms conventional transmission schemes such as Non-orthogonal Multiple Access (NOMA) in terms of MMF.
\end{abstract}

\section{Introduction}
Rate-Splitting Multiple Access (RSMA) has emerged as a promising and robust multiple access and interference management strategy for wireless networks\cite{mao2022rate}, \cite{clerckx2023primer}. It holds significant potential for addressing challenges such as higher throughput and ultra reliability in next-generation communication systems. The fundamental concept of RSMA involves splitting the user messages into multiple parts, and employing Successive Interference Cancellation (SIC) at all receivers to sequentially recover the different parts. 

In the downlink, RSMA stands out by bridging and surpassing Space Division Multiple Access (SDMA) which fully treats interference as noise, and Non-orthogonal Multiple Access (NOMA) which fully decodes interference from both spectral and energy effeciency perspectives under perfect and imperfect Channel State Information at the Transmitter (CSIT)\cite{mao2018rate,clerckx2023primer,mao2019rate,mao2022rate}. In addition to the spectral and energy efficiency, Max Minimum Fairness (MMF) is identified as another crucial criterion. In \cite{joudeh2017rate} and \cite{xu2022max}, the authors have demonstrated that RSMA yields significant user fairness enhancement in the multi-user downlink communications with both Infinite Blocklength (IFBL) and Finite Blocklength (FBL). Moreover, it has been substantiated in \cite{yalcin2021max} that RSMA exhibits superior MMF compared to NOMA in a downlink Multi-User Multiple-Input Multiple-Output (MU-MIMO) system. 

In addition to its effectiveness in the downlink, various recent studies have showcased the superior performance of RSMA in the uplink. \cite{rimoldi1996rate} has established that the capacity region of a $K$-user Gaussian Multiple Access Channel (MAC) can be attained through uplink RSMA by treating it as $2K-1$ virtual point-to-point channels through message splitting. In contrast, NOMA necessitates the assistance of time-sharing to achieve the capacity region. \cite{zeng2019ensuring} has comfirmed that RSMA exhibits a superior worst-case transmission rate among users and also reduces transmission latency in uplink Single-Input Multiple-Output (SIMO) systems when compared to NOMA. 
However, all the aforementioned works on uplink RSMA are limited to the ideal assumption of IFBL. 

A key domain where uplink multiple-access schemes find significant application is in Ultra-Reliable Low-Latency Communications (URLLC). Low-latency demands of URLLC necessitate systems to function with short blocklength, emphasizing the importance of analyzing system performance using FBL codes. In the groundbreaking work \cite{polyanskiy2010channel}, the authors have established the achievable rates with FBL codes for a given blocklength and error probability. This work has paved the way for theoretical performance analysis utilizing FBL codes. \cite{xu2022rate, liu2022network} have demonstrated that uplink RSMA can provide a more reliable service with a lower error probability and a higher arrival data rate compared to NOMA in URLLC scenario. However, these works are limited to the two-user case and do not address the user fairness issue.

Motivated by the benefits offered by RSMA in terms of error probability, latency and throughput, alongside the identified research gaps, we assess in this letter the benefits of $K$-user uplink RSMA with the objective of enhancing the MMF under FBL constraints. The contributions of this paper are summarized as follows:
\begin{itemize}
\item This paper establishes a general $K$-user uplink RSMA system model, wherein an arbitrary group of users performs message splitting and arbitrary decoding order is applied at the receiver. Additionally, we introduce a low-complexity algorithm for designing the decoding order at the receiver side.
\item Based on the established model, we formulate a MMF problem under FBL constraints to design the power allocation among the splitting messages for each user. Then, we propose a successive convex approximation (SCA) based method to address the formulated problem. 
\item Numerical results demonstrate that by in the $K$-user MAC, FBL RSMA achieves superior user fairness compared to FBL NOMA by splitting the message of only one user into two parts. This contrasts to the convention of splitting the messages of $K-1$ users as specified in \cite{rimoldi1996rate}. Furthermore, FBL RSMA can even outperform IFBL NOMA by increasing the number of splitting users.
\end{itemize}

\textit{Organizations}:  The system model of uplink RSMA is specified in Section \ref{2}. The problem is formulated and optimized in Section \ref{3}. Section \ref{4} presents the numerical results and we conclude the letter in Section \ref{5}.

\emph{Notations}: Italic, bold lower-case and calligraphic letters denote scalars, vectors and sets respectively. $|\cdot|$ denotes the absolute value if the argument is a scalar or the cardinality if the argument is a set. $\cdot\setminus\cdot$ denotes the difference between two sets if the arguments are sets.

\section{System Model}\label{2}
We first consider a $K$-user Single-Input Single-Output (SISO) MAC with perfect CSIT and perfect Channel State Information at Receiver (CSIR). The $K$ single-antenna users are indexed by $\mathcal{K}=\{1, ..., K\}$, and they simultaneously send information to a Base Station (BS) with a single antenna. In uplink RSMA, only partial users need to split their messages into two parts and others remain non-splitting. Thus, we use $\mathcal{J}\triangleq\{j_{1}, ..., j_{|\mathcal{J}|}\}$ to denote the splitting users and $\mathcal{U}=\mathcal{K}\setminus\mathcal{J}\triangleq\{u_{1}, ..., u_{|\mathcal{U}|}\}$ to denote the non-splitting users. The message of user-$j_{l}$ in $\mathcal{J}$ are split into two parts as $W_{j_{l}}=\{W_{j_{l},1}, W_{j_{l},2}\}$, then they are independently encoded into $\textbf{s}_{j_{l}}=[s_{j_{l},1}, s_{j_{l},2}], j_{l}\in\mathcal{J}, l\in\{1, ..., |\mathcal{J}|\}$. The message $W_{u_q}$ of user-$u_{q}$ in $\mathcal{U}$, $W_{u_{q}}$ is directly encoded into $s_{u_{q}}, u_{q}\in\mathcal{U}, q\in\{1, ..., |\mathcal{U}|\}$, resulting in a total of $2|\mathcal{J}|+|\mathcal{U}|$ streams denoted as $\textbf{s}=[\textbf{s}^{\mathcal{J}}, \textbf{s}^{\mathcal{U}}]$, where $\textbf{s}^{\mathcal{J}}=[s_{j_{1},1}, s_{j_{1},2}, ..., s_{j_{|\mathcal{J}|},1}, s_{j_{|\mathcal{J}|},2}]$ and $\textbf{s}^{\mathcal{U}}=[s_{u_{1}}, ..., s_{u_{|\mathcal{U}|}}]$ represent the streams for the splitting and non-splitting users, respectively. Let $P_{k}, k\in\mathcal{K}$ denotes the transmit power of each user. The received signal at the BS is written as 
\begin{equation}
    y = \underbrace{\sum_{j_{l}\in\mathcal{J}}h_{j_{l}}\sum_{d=1}^{2}\sqrt{P_{j_{l},d}}s_{j_{l},d}}_{\text{splitting users}}+\underbrace{\sum_{u_{q}\in\mathcal{U}}h_{u_{q}}\sqrt{P_{u_{q}}}s_{u_{q}}}_{\text{non-splitting users}}+n, \label{eq:1}
\end{equation}
where $P_{j_{l},d}$ is the transmit power of $s_{j_{l},d}, d\in\{1,2\}$ from the splitting user-$j_{l}$ and $\sum_{d=1}^{2}P_{j_{i},d}\leq P_{j_{l}}, j_{l}\in\mathcal{J}$. $n \sim\mathcal{CN}(0,\sigma_n^2)$ is the Additive White Gaussian Noise (AWGN) at the BS. Without loss of generality, we assume the noise variance is one, i.e., $\sigma_{n}^{2}=1$. Then, the transmit SNR is numerically equal to the transmit power.

The user signals are superimposed at the BS. In order to decode all data streams, SIC is applied. There are $(2|\mathcal{J}|+|\mathcal{U}|)!$ possible decoding orders 
The optimal decoding order can be determined through exhaustive search, but its complexity grows exponentially with the number of users and data streams. To address this issue, we here propose a low-complexity decoding order scheme inspired by \cite{liu2020rate}, which involves separating the parts of splitting messages from one user and arranging non-splitting users by the descending order of the channel gain. Assume that $|h_{u_{j_{1}}}|\geq|h_{u_{j_{2}}}|\geq...\geq|h_{u_{j_{|\mathcal{J}|}}}|$ and $|h_{u_{q_{1}}}|\geq|h_{u_{q_{2}}}|\geq...\geq|h_{u_{q_{|\mathcal{U}|}}}|$, the proposed decoding order is $s_{j_{1},1}\to ...\to s_{j_{|\mathcal{J}|},1}\to s_{u_{1}}\to ...\to s_{u_{|\mathcal{U}|}}\to s_{j_{1},2}\to, ..., \to s_{j_{|\mathcal{J}|},2}$.
We map this decoding order into $\bm{\pi}=[s_{\bm{\pi}_{1}}, ..., s_{\bm{\pi}_{2|\mathcal{J}|+|\mathcal{U}|}}]$. These $2|\mathcal{J}|+|\mathcal{U}|$ streams are indexed by $\mathcal{G}=\{1, ..., 2|\mathcal{J}|+|\mathcal{U}|\}$. The $g^{th}$  stream 
$s_{g}, g\in\mathcal{G}$ is decoded at the BS by treating the remaining $g+1^{th}$ to $2|\mathcal{J}|+|\mathcal{U}|^{th}$ streams as noise. Once $s_{\bm{\pi}_{g}}$ is successfully decoded, $\widehat{W}_{\bm{\pi}_{g}}$ is obtained, reconstructed and subtracted from the current received signal. Thus, the Signal-to-Interference-Noise Ratio (SINR) $\gamma_{\bm{\pi}_{g}}$ of stream $s_{\bm{\pi}_{g}}$ is 
\begin{equation}
    \gamma_{\bm{\pi}_{g}} =\frac{P_{\bm{\pi}_{g}}|h_{\bm{\pi}_{g}}|^2}{\sum_{i=g+1}^{2|\mathcal{J}|+|\mathcal{U}|}P_{\bm{\pi}_{i}}|h_{\bm{\pi}_{i}}|^2+1}. \label{eq:2}
\end{equation}
Once all streams are successfully decoded, the estimated messages for split users in $\mathcal{U}$ are obtained by combining the corresponding two split messages, e.g., the message $\widehat{W}_{j_{1}}$ for user-$j_{1}$ is obtained by combining $\widehat{W}_{\bm{\pi}_{1}}$ and $\widehat{W}_{\bm{\pi}_{1+K}}$.
Then, the FBL achievable rate expression of stream $s_{\bm{\pi}_{g}}, g\in\mathcal{G}$ is written as \cite{polyanskiy2010channel} 
\begin{equation}\label{eq:3}
    r_{\bm{\pi}_{g}}=\log_2(1+\gamma_{\bm{\pi}_{g}})-\frac{B}{N}\sqrt{V_{\gamma_{\bm{\pi}_{g}}}},
\end{equation}
where $B=Q^{-1}(\epsilon)\log_2(e)$, $\epsilon$ is the error probability and $Q$ is the Q-function. \footnote{Q-function is the tail distribution function of the standard normal distribution. Normally, Q-function is defined as: $Q(x)=\frac{1}{\sqrt{2\pi}} \int_x^{\infty}e^{-\frac{u^2}{2}}du.$} $\gamma$ is the SINR of the stream, $N$ is the blocklength and 
$V = \biggl(1-(1+\gamma_{\bm{\pi}_{g}})^{-2}\biggl)$ 
is the channel dispersion\cite{schiessl2020noma}. The total achievable rates of splitting users can be calculated by summing the transmission rates for the two corresponding split streams, e.g., the rate of user-$j_{1}$ in $\mathcal{J}$ equals $r_{\bm{\pi}_{1}}+r_{\bm{\pi}_{1+K}}$ 
and the rates of non-splitting users such as user-$u_{1}$ in $\mathcal{U}$ equals $r_{\bm{\pi}_{|\mathcal{J}|+1}}$ individually.

\textit{Remark 1}: The system model of NOMA is a particular instance of the model we specified in this section. By grouping all users into the non-splitting user set, RSMA simplifies to NOMA. Then, the message $W_{k}$ of user-$k, k\in\mathcal{K}$ is encoded into $s_{k}$. The received signal at the BS is expressed as $y = \sum_{k\in\mathcal{K}}h_{k}\sqrt{P_{k}}s_{k}+n$. The decoding order is arranged by the descending order of the channel gain. Accordingly, the rate of each user can be calculated by eq. (\ref{eq:3}).

\section{Problem Formulation and Optimization}\label{3}
In this work, our objective is to maximize the minimum user rate among users by optimizing the power allocation of the splitting messages. The general max-min rate problem for the $K$-user uplink SISO RSMA system is formulated as
\begin{subequations}
    \begin{align}
        \max_{\textbf{p}_{\bm{\pi}}} \quad \min{R_{k}}\\
        \mbox{s.t.} \quad
        & P_{j_{l},1}+P_{j_{l},2}\leq P_{t}, j_{l}\in \mathcal{J} \\
        & P_{u_{q}}\leq P_{t}, u_{q}\in \mathcal{U},
    \end{align} 
    \label{Prob:4}
\end{subequations}
where $R_{k}$ is the rate of user-$k$ in $\mathcal{K}$ and $\textbf{p}_{\bm{\pi}}=[P_{j_{1},1}, ..., P_{j_{|\mathcal{J}|},1}, P_{u_{1}}, ..., P_{u_{|\mathcal{U}|}}, P_{j_{1},2}, ..., P_{j_{|\mathcal{J}|},2}]$ is the power allocation of each stream corresponding to the decoding order $\bm{\pi}$. 
We assume the maximum transmit power $P_{t}$ is equal for all users. 

Since the rate and SINR equations are not convex, we proposed a SCA-based algorithm to solve this problem by introducing slack variables and successively approximating the non-convex expressions by their first-order Taylor approximation. Specifically, a slack variable $t$ is first introduced, according to the decoding order $\bm{\pi}$ we proposed, Problem (\ref{Prob:4}) is transformed into 
\begin{subequations}
    \begin{align}
        \max_{\textbf{p}_{\bm{\pi}}, t} \quad t \\
        \mbox{s.t.} \quad
        & r_{l}+r_{l+K}\geq t, l\in\{1, ..., |\mathcal{J}|\} \label{eq:5b}\\
        & r_{o}\geq t, o\in\{|\mathcal{J}|, ..., K\} \label{eq:5c}\\
        & P_{l}+P_{l+K}\leq P_{t}, l\in\{1, ..., |\mathcal{J}|\} \label{eq:5d} \\
        & P_{o}\leq P_{t}, o\in\{|\mathcal{J}|, ..., K\}, \label{eq:5e}
    \end{align} 
    \label{Prob:5}
\end{subequations} 
where in the decoding order $\bm{\pi}$, the messages $s_{j_{l},1}$ and $s_{j_{l},2}$ of each splitting users  are mapped to $s_{l}$ and $s_{l+K}, l\in\{1, ..., |\mathcal{J}|\}$, respectively. Moreover, the message of non-splitting users $s_{u_{q}}$ is mapped to $s_{o}, o\in\{|\mathcal{J}|, ..., K\}$. 
Constraints \eqref{eq:5b} and \eqref{eq:5c} are the rate constraints for the splitting and non-splitting users, respectively. 
We further introduce slack variables $\bm{\rho}=[\rho_{1}, ..., \rho_{2|J|+|\mathcal{U}|}]$ and transform Problem (\ref{Prob:5}) into
\begin{subequations}
    \begin{align}
        \max_{\textbf{p}_{\bm{\pi}}, \bm{\rho}, t} \quad t \\
        \mbox{s.t.} \quad
        & \log_2(1+\rho_{l})-\frac{B}{\sqrt{N}}\sqrt{V_{\rho_{l}}}+\log_2(1+\rho_{l+K}) \nonumber \\
        & -\frac{B}{\sqrt{N}}\sqrt{V_{\rho_{l+K}}}\geq t, l\in(1, ..., |\mathcal{J}|) \label{eq:6b} \\
        & \log_2(1+\rho_{o})-\frac{B}{\sqrt{N}}\sqrt{V_{\rho_{o}}}\geq t, o\in(|\mathcal{J}|, ..., K) \label{eq:6c} \\
        & \frac{P_{\bm{\pi}_{g}}|h_{\bm{\pi}_{g}}|^2}{\sum_{i=g+1}^{2|\mathcal{J}|+|\mathcal{U}|}P_{\bm{\pi}_{i}}|h_{\bm{\pi}_{i}}|^2+1}\geq\rho_{g}, g\in\mathcal{G} \label{eq:6d} \\
        & \eqref{eq:5d}, \eqref{eq:5e}. \nonumber
    \end{align} 
    \label{Prob:6}
\end{subequations}
Problem (\ref{Prob:6}) remains non-convex due to the contraints \eqref{eq:6b}-\eqref{eq:6d}. Next, we approximate the non-convex terms $\sqrt{V_{\rho_{l}}}$, $\sqrt{V_{\rho_{l+K}}}$ and $\sqrt{V_{\rho_{o}}}$ by their first-order Taylor approximation around the point $(\bm{\rho}^{[n]})$ at iteration $n$, which are expressed as \cite{xu2022max} 
\begin{subequations}
    \begin{align}
        \sqrt{V_{\rho_{l}}} & \geq \sqrt{1-(1+\rho_{l}^{[n]})^{-2}}+(\rho_{l}-\rho_{l}^{[n]}) \nonumber \\
        & (1+\rho_{l}^{[n]})^{-3}(1-(1+\rho_{l}^{[n]})^{-2})^{-\frac{1}{2}}\triangleq \Phi_{l}^{[n]}\\
        \sqrt{V_{\rho_{l+K}}} & \geq \sqrt{1-(1+\rho_{l+K}^{[n]})^{-2}}+(\rho_{l+K}-\rho_{l+K}^{[n]}) \nonumber \\
        & (1+\rho_{l+K}^{[n]})^{-3}(1-(1+\rho_{l+K}^{[n]})^{-2})^{-\frac{1}{2}}\triangleq \Phi_{l+K}^{[n]} \\
        \sqrt{V_{\rho_{o}}} & \geq \sqrt{1-(1+\rho_{o}^{[n]})^{-2}}+(\rho_{o}-\rho_{o}^{[n]}) \nonumber \\
        &(1+\rho_{o}^{[n]})^{-3}(1-(1+\rho_{o}^{[n]})^{-2})^{-\frac{1}{2}}\triangleq \Phi_{o}^{[n]},
    \end{align}
\end{subequations}
For constraint \eqref{eq:6d}, we approximate it around the point $(\textbf{p}_{\bm{\pi}}^{[n]}, \bm{\rho}^{[n]})$ at iteration $n$  as 
\begin{equation} \label{eq:8}
    \begin{split}
        & \sum_{i=g+1}^{2|\mathcal{J}|+|\mathcal{U}|}P_{\bm{\pi}_{i}}|h_{\bm{\pi}_{i}}|^2+1 \\ 
        & -\frac{P_{\bm{\pi}_{g}}|h_{\bm{\pi}_{g}}|^2}{\rho_{g}^{[n]}}+(\rho_{g}-\rho_{g}^{[n]})\frac{P_{\bm{\pi}_{g}}^{[n]}|h_{\bm{\pi}_{g}}|^2}{(\rho_{g}^{[n]})^2}\leq 0.        
    \end{split}
\end{equation}
Based on the above approximation methods, the original non-convex problem is transformed into a convex one and can be solved using the SCA method. At iteration $n$, based on the optimal solution 
$(\textbf{p}_{\bm{\pi}}^{[n]}, \bm{\rho}^{[n]})$ obtained from the previous iteration $n-1$, we solve the following problem:
\begin{subequations} \label{Prob:9}
    \begin{align}
        \min_{\textbf{p}_{\bm{\pi}},\bm{\rho},t} \quad & t\\
        \mbox{s.t.}\quad
        & \log_2(1+\rho_{l})-\frac{B}{\sqrt{N}}\Phi_{l}^{[n]}+\log_2(1+\rho_{l+K}) \nonumber\\
        & -\frac{B}{\sqrt{N}}\Phi_{l+K}^{[n]}\geq t, l\in(1, ..., |\mathcal{J}|) \\
        & \log_2(1+\rho_{o})-\frac{B}{\sqrt{N}}\Phi_{o}^{[n]}\geq t, o\in(|\mathcal{J}|, ..., K),  \\
        & \eqref{eq:8}, \eqref{eq:5d}, \eqref{eq:5e}. \nonumber
    \end{align}
\end{subequations}
The SCA-based power allocation algorithm is outlined in Algorithm 1. $\tau$ represents the tolerance of convergence. 
\begin{algorithm}[t] \label{Al:1} 
\caption{Proposed SCA-based algorithm} 
\LinesNumbered
\SetKwInput{kwInit}{Initialize}
\kwInit{$n\gets0,t^{[n]}\gets0,\textbf{P}_{\bm{\pi}}^{[n]},\bm{\rho}^{[n]}$;}
\Repeat{$|t^{[n]}-t^{[n-1]}|\leq \tau$}
 {$n\gets n+1$; \\
 Solve problem (\ref{Prob:9}) using $\textbf{P}_{\bm{\pi}}^{[n-1]}, \bm{\rho}^{[n-1]}$ and find optimal $t^*$, $\textbf{P}_{\bm{\pi}}^*,\bm{\rho^*}$; \\
 Update $t^{[n]}\gets t^*, \textbf{P}_{\bm{\pi}}^{[n]}\gets \textbf{P}_{\bm{\pi}}^*, \bm{\rho^{[n]}\gets\bm{\rho^*}}$;}
\end{algorithm}
At each SCA iteration, the approximated problem is solved. 

\textit{Convergence and Complexity  Analysis}: Since the solution of Problem (\ref{Prob:9}) at iteration $n-1$ is a feasible point of Problem (\ref{Prob:9}) at iteration $n$, and the objective function $t$ is monotonically increasing and it is bounded by the transmit power constraints \eqref{eq:5d} and \eqref{eq:5e}. 
The total number of SCA iterations required for convergence is approximated as $\mathcal{O}\left(\log(\tau^{-1})\right)$. The worst-case of computational complexity of the SCA is $\mathcal{O}\left(\log(\tau^{-1})[K]^{3.5}\right)$. Hence, the computational complexity for the proposed algorithm is $\mathcal{O}\left(\log(\tau^{-1})[K]^{3.5}\right)$.

\section{Numerical Results}\label{4}
This section investigates the MMF of the proposed scheme and compares it with conventional transmission schemes. The tolerance of the algorithm is set to $\tau=10^{-3}$. We perform simulations for Rayleigh Fading channel with the number of channel realizations of 100. Here, we compare the performance of RSMA, NOMA and Treat-Interference-as-Noise (TIN). 
\begin{itemize}
\item NOMA: This is a special case of RSMA. As specified in Remark 1, none of the users split their messages in NOMA. The BS sequentially decodes the user messages based on SIC. This has been studied in \cite{schiessl2020noma,zhang2023max}.
\item TIN: This is a classical method where none of the users split their messages, and the BS decodes the message of each user by treating the messages of all other users as interference. This has been extensively studied in \cite{shang2009new}.
\end{itemize}

\begin{figure}[t]
    \centering
    \includegraphics[width=3in]{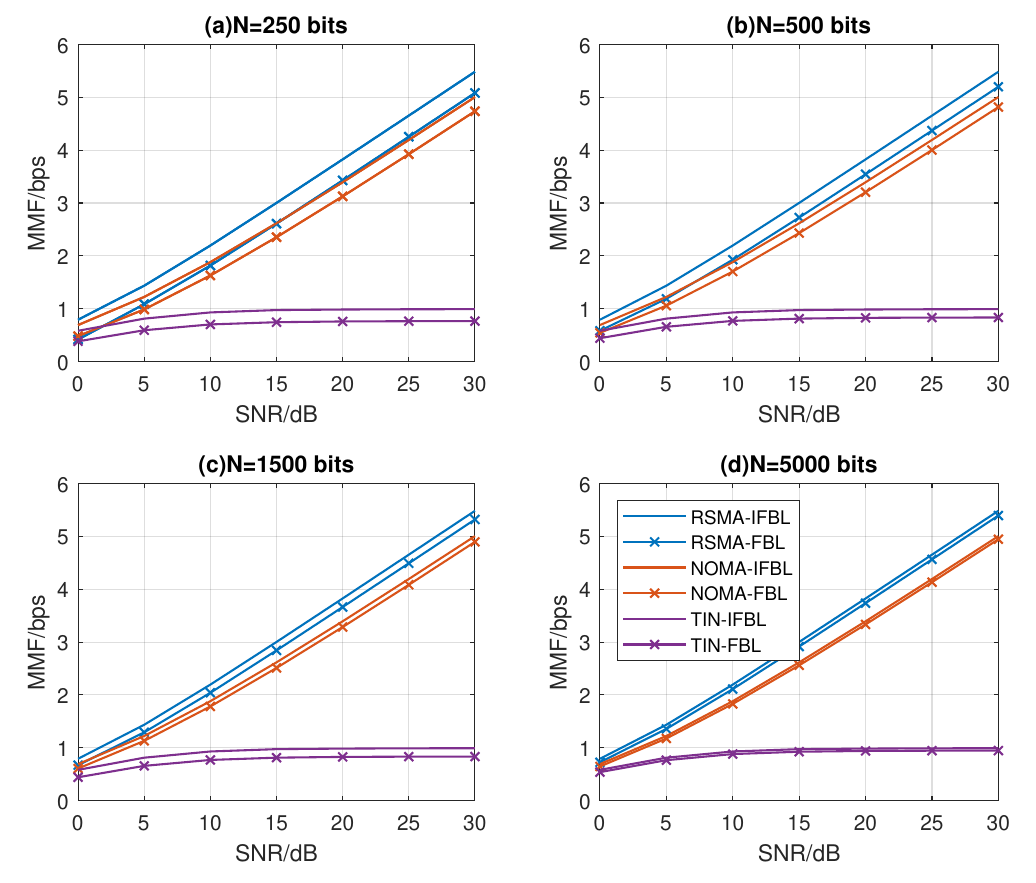}
    \caption{The MMF performance of RSMA, NOMA and TIN versus transmit SNR with four different blocklengths. K=2.}
    \label{Fig.1}
\end{figure} 

\begin{figure}[t]
    \centering
    \includegraphics[width=3in]{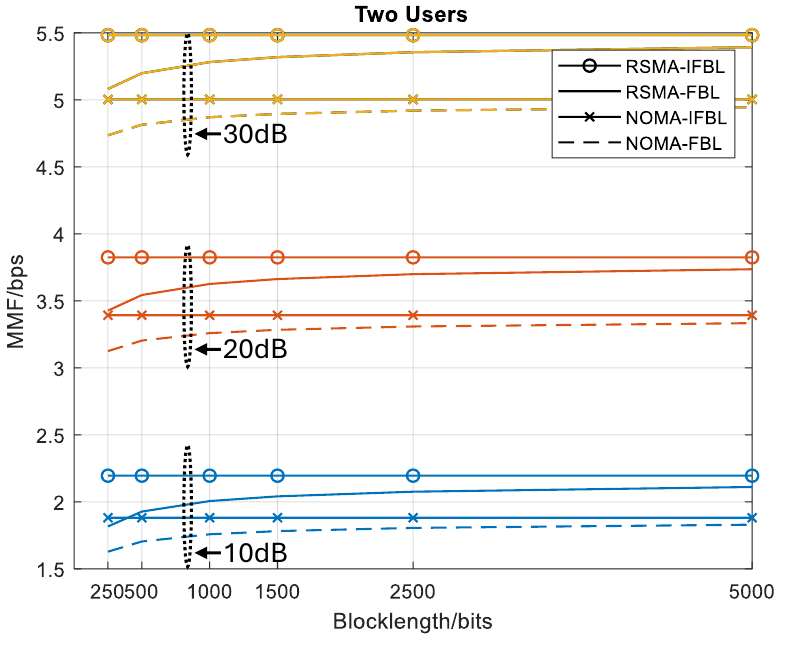}
    \caption{The MMF performance of RSMA and NOMA versus blocklengths with three different transmit SNR. K=2.}
    \label{Fig.2}
\end{figure} 

Fig. \ref{Fig.1} compares the MMF of RSMA, NOMA, and TIN with different blocklengths versus transmit SNR when the number of users $K$ is two. We only consider splitting any one user's message since \cite{liu2020rate} demonstrated the result would be the same with splitting both two users. Four different blocklengths $N$ of $250, 500, 1500$, and $5000$ bits are consiered. In Fig. \ref{Fig.1}, it is evident that RSMA demonstrates superior performance compared to NOMA and TIN. Fig. \ref{Fig.1} (a) shows that even with a very short blocklength of $N=250$ bits, the MMF of FBL RSMA outperforms IFBL NOMA in the high SNR regime. When blocklength increases, as shown in Fig. \ref{Fig.1} (c), with $N=1500$ bits RSMA performs better than IFBL NOMA for all SNR. TIN constantly performs badly. Therefore, we ignore TIN in the following analysis.

\begin{figure}[t]
    \centering
    \includegraphics[width=3in]{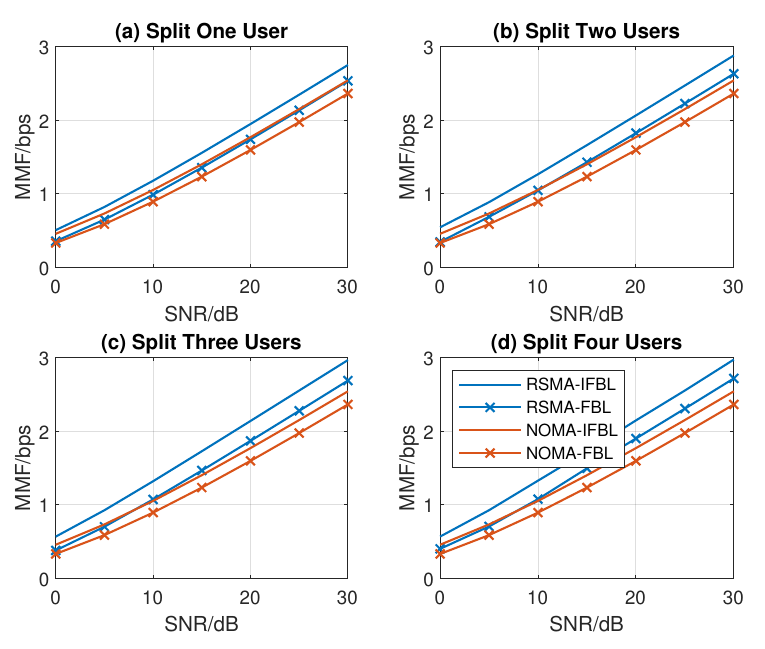}
    \caption{The MMF performance versus transmit SNR of RSMA and NOMA with different numbers of splitting users. K=4. Blocklength=250 bits.}
    \label{Fig.3}
\end{figure} 

\begin{figure}[t]
    \centering
    \includegraphics[width=3in]{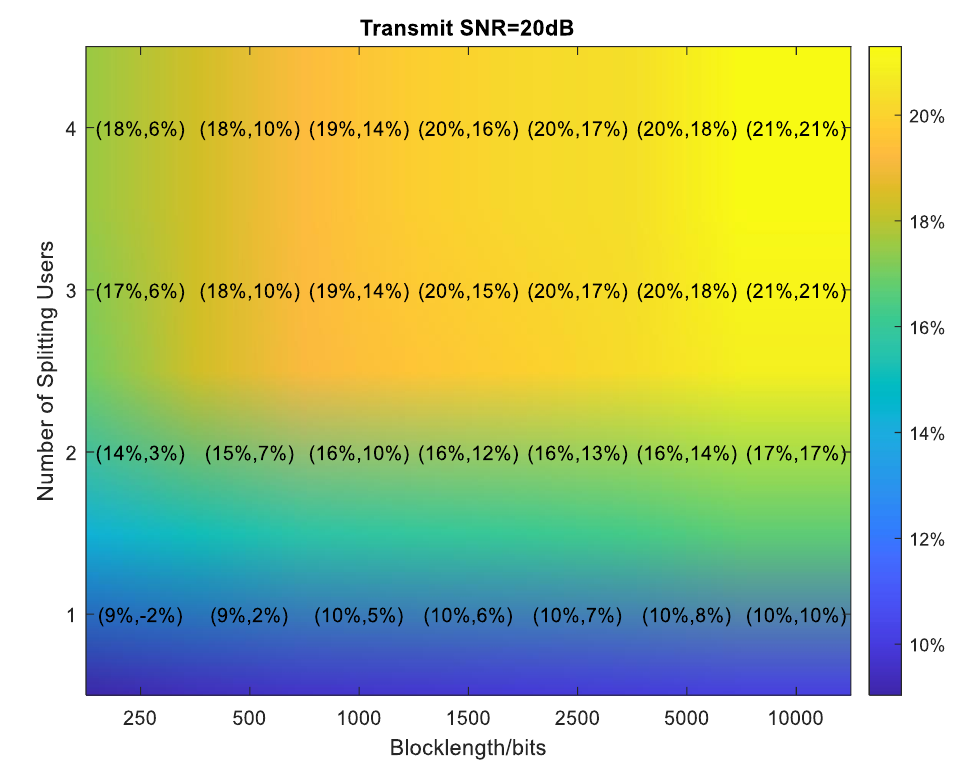}
    \caption{Relative MMF rate gain of RSMA compared to NOMA. K=4. Transmit SNR=20dB.}
    \label{Fig.4}
\end{figure} 

Fig. \ref{Fig.2} shows the trend of MMF as blocklength increases. We can observe that FBL RSMA is always better than FBL NOMA. With a transmit SNR of 10 dB, the MMF of RSMA with blocklength of 250 bits is slightly lower than IFBL NOMA. When it increases to 20 dB and 30 dB, the MMF rate of FBL RSMA is better than IFBL NOMA. 
Since NOMA is strictly constrained by the decoding order, the SINR of the first decoded stream is influenced by the transmit power of the second decoded stream, then the first decoded user would transmit with the full power while the second would transmit with less power to balance the SINRs between two streams. For RSMA, the decoding order is $s_{1,1}\to s_{2}\to s_{1,2}$ if user-$1$ splits its message, \footnote{If user-2 splits its message, the decoding order would be $s_{2,1}\to s_{1}\to s_{2,2}$.} and the result shows $s_{1,1}$ is transmitted with around 93\% of the transmit power of user-$1$ leaving only 7\% for $s_{1,2}$ while user-$2$ now transmits with full power. Thus, although the SINR of $s_{1,1}$ would be smaller than the first decoded stream in NOMA, $s_{1,2}$ is interference-free which can compensate for the loss.

In the two-user case, we only consider one splitting 
user. Now we investigate the impact of the number of splitting users in the four-user case. In Fig. \ref{Fig.3}, we compare the MMF of NOMA to the following schemes: RSMA a) with one; b) with two; c) with three; d) with four splitting users, respectively. Blocklength is $250$ bits and the results of RSMA and NOMA with IFBL are also included. The MMF of IFBL NOMA is marginally better than FBL RSMA when SNR is low and only one splitting user is considered in Fig. \ref{Fig.3} (a). We observe that RSMA with more splitting users can yield higher MMF in Fig. \ref{Fig.3}. The relative gain of FBL RSMA compared to FBL NOMA is illustrated in Fig. \ref{Fig.4} and it is defined by
\begin{equation}
    \frac{R^{\text{RSMA}}_{\text{MMF}}-R^{\text{NOMA}}_{\text{MMF}}}{R^{\text{NOMA}}_{\text{MMF}}}\times100\%.
\end{equation}
The percentages in parentheses represent MMF rate gains of FBL RSMA over FBL NOMA (with the same blocklength) and IFBL NOMA. FBL RSMA can have some gains over IFBL NOMA in most of the cases. With FBL, increasing the number of splitting users from 1 to 3 would bring a gain of around 8\%.

\begin{figure}[t]
    \centering
    \includegraphics[width=3in]{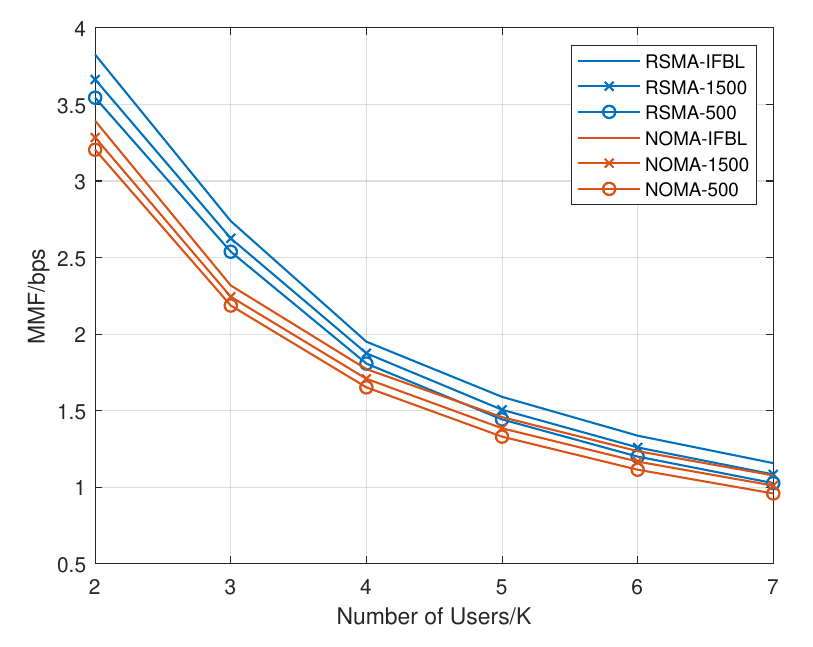}
    \caption{The MMF versus the number of users K of RSNMA and NOMA. Transmit SNR=20 dB.}
    \label{Fig.5}
\end{figure}

Fig. \ref{Fig.5} displays the MMF with different blocklengths versus the number of users $K$ with a transmit SNR of $20$ dB. Only one splitting user is considered in this figure. FBL RSMA with blocklength of $500$ bits outperforms FBL NOMA with both $500$ bits and $1500$ bits. When blocklength increases to $1500$ bits, FBL RSMA is even better than IFBL NOMA. We can see that with only one splitting user, FBL RSMA can still outperform IFBL NOMA at the sacrifice of only one additional SIC layer at the receiver. 

\section{Conclusion}\label{5}
The purpose of this letter is to investigate if RSMA can enhance the max-min user rate performance of a general $K$-user SISO MAC with FBL. Specifically, we propose a low-complexity decoding order method at the receiver side and then propose a SCA-based method to solve the MMF power allocation problem. Extensive numerical results demonstrate that, for a given blocklength, FBL RSMA can achieve a higher MMF than FBL NOMA. With the increase of blocklength and number of splitting users, FBL RSMA can perform better than IFBL NOMA. Even with one splitting user, FBL RSMA can outperform IFBL NOMA SISO MAC. The MMF decreases as the number of users $K$ rises, but RSMA consistently outperforms NOMA. In summary, RSMA consistently achieves higher MMF than NOMA regardless of FBL or IFBL and the performance of RSMA can be improved by increasing the number of splitting users, but at the sacrifice of a higher hardware and computational complexities. Moreover, RSMA has the capability to reduce the blocklength compared to NOMA while achieving the same MMF, thereby effectively reducing latency.

\bibliographystyle{IEEEtran}

\bibliography{ref}

\begin{thebibliography}{10}
\providecommand{\url}[1]{#1}
\csname url@samestyle\endcsname
\providecommand{\newblock}{\relax}
\providecommand{\bibinfo}[2]{#2}
\providecommand{\BIBentrySTDinterwordspacing}{\spaceskip=0pt\relax}
\providecommand{\BIBentryALTinterwordstretchfactor}{4}
\providecommand{\BIBentryALTinterwordspacing}{\spaceskip=\fontdimen2\font plus
\BIBentryALTinterwordstretchfactor\fontdimen3\font minus \fontdimen4\font\relax}
\providecommand{\BIBforeignlanguage}[2]{{%
\expandafter\ifx\csname l@#1\endcsname\relax
\typeout{** WARNING: IEEEtran.bst: No hyphenation pattern has been}%
\typeout{** loaded for the language `#1'. Using the pattern for}%
\typeout{** the default language instead.}%
\else
\language=\csname l@#1\endcsname
\fi
#2}}
\providecommand{\BIBdecl}{\relax}
\BIBdecl

\bibitem{mao2022rate}
Y.~Mao, O.~Dizdar, B.~Clerckx, R.~Schober, P.~Popovski, and H.~V. Poor, ``Rate-splitting multiple access: Fundamentals, survey, and future research trends,'' \emph{IEEE Communications Surveys \& Tutorials}, 2022.

\bibitem{clerckx2023primer}
B.~Clerckx, Y.~Mao, E.~A. Jorswieck, J.~Yuan, D.~J. Love, E.~Erkip, and D.~Niyato, ``A primer on rate-splitting multiple access: Tutorial, myths, and frequently asked questions,'' \emph{IEEE Journal on Selected Areas in Communications}, 2023.

\bibitem{mao2018rate}
Y.~Mao, B.~Clerckx, and V.~O. Li, ``Rate-splitting multiple access for downlink communication systems: bridging, generalizing, and outperforming {SDMA} and {NOMA},'' \emph{EURASIP journal on wireless communications and networking}, vol. 2018, pp. 1--54, 2018.

\bibitem{mao2019rate}
------, ``Rate-splitting for multi-antenna non-orthogonal unicast and multicast transmission: Spectral and energy efficiency analysis,'' \emph{IEEE Transactions on Communications}, vol.~67, no.~12, pp. 8754--8770, 2019.

\bibitem{joudeh2017rate}
H.~Joudeh and B.~Clerckx, ``Rate-splitting for max-min fair multigroup multicast beamforming in overloaded systems,'' \emph{IEEE Transactions on Wireless Communications}, vol.~16, no.~11, pp. 7276--7289, 2017.

\bibitem{xu2022max}
Y.~Xu, Y.~Mao, O.~Dizdar, and B.~Clerckx, ``Max-min fairness of rate-splitting multiple access with finite blocklength communications,'' \emph{IEEE Transactions on Vehicular Technology}, 2022.

\bibitem{yalcin2021max}
A.~Z. Yalcin, M.~K. Cetin, and M.~Yuksel, ``Max-min fair precoder design and power allocation for {MU-MIMO NOMA},'' \emph{IEEE Transactions on Vehicular Technology}, vol.~70, no.~6, pp. 6217--6221, 2021.

\bibitem{rimoldi1996rate}
B.~Rimoldi and R.~Urbanke, ``A rate-splitting approach to the gaussian multiple-access channel,'' \emph{IEEE Transactions on Information Theory}, vol.~42, no.~2, pp. 364--375, 1996.

\bibitem{zeng2019ensuring}
J.~Zeng, T.~Lv, W.~Ni, R.~P. Liu, N.~C. Beaulieu, and Y.~J. Guo, ``Ensuring max--min fairness of {UL SIMO-NOMA}: A rate splitting approach,'' \emph{IEEE Transactions on Vehicular Technology}, vol.~68, no.~11, pp. 11\,080--11\,093, 2019.

\bibitem{polyanskiy2010channel}
Y.~Polyanskiy, H.~V. Poor, and S.~Verd{\'u}, ``Channel coding rate in the finite blocklength regime,'' \emph{IEEE Transactions on Information Theory}, vol.~56, no.~5, pp. 2307--2359, 2010.

\bibitem{xu2022rate}
J.~Xu, O.~Dizdar, and B.~Clerckx, ``Rate-splitting multiple access for short-packet uplink communications: A finite blocklength analysis,'' \emph{IEEE Communications Letters}, 2022.

\bibitem{liu2022network}
Y.~Liu, B.~Clerckx, and P.~Popovski, ``Network slicing for {eMBB}, {URLLC}, and {mMTC}: An uplink rate-splitting multiple access approach,'' \emph{arXiv preprint arXiv:2208.10841}, 2022.

\bibitem{liu2020rate}
H.~Liu, T.~A. Tsiftsis, K.~J. Kim, K.~S. Kwak, and H.~V. Poor, ``Rate splitting for uplink {NOMA} with enhanced fairness and outage performance,'' \emph{IEEE Transactions on Wireless Communications}, vol.~19, no.~7, pp. 4657--4670, 2020.

\bibitem{schiessl2020noma}
S.~Schiessl, M.~Skoglund, and J.~Gross, ``{NOMA} in the uplink: Delay analysis with imperfect {CSI} and finite-length coding,'' \emph{IEEE Transactions on Wireless Communications}, vol.~19, no.~6, pp. 3879--3893, 2020.

\bibitem{zhang2023max}
Y.~Zhang, T.~Zhong, Y.~Wang, J.~Wang, K.~Zheng, and X.~You, ``Max-min fairness for uplink {NOMA} systems with finite blocklength,'' \emph{IEEE Transactions on Vehicular Technology}, 2023.

\bibitem{shang2009new}
X.~Shang, G.~Kramer, and B.~Chen, ``A new outer bound and the noisy-interference sum--rate capacity for gaussian interference channels,'' \emph{IEEE Transactions on Information theory}, vol.~55, no.~2, pp. 689--699, 2009.

\end{thebibliography}


\end{document}